\documentclass[conference]{IEEEtran}
\ifCLASSINFOpdf
  \usepackage[pdftex]{graphicx}
\else
\fi

\usepackage{comment}
\usepackage{enumitem}
\usepackage{graphicx}
\begin{document}
\title{Driver State and Behavior Detection Through Smart Wearables}

%Title: 
% Classifying Driving Behavior, Events, and Environmental Conditions in the Wild through Wearable

\author{\IEEEauthorblockN{Arash Tavakoli}
\IEEEauthorblockA{Department of Engineering\\Systems and Environment\\
University of Virginia\\
Charlottesville,\\ Virginia, 22903\\
Email: at5cf@virginia.edu}
\and
\IEEEauthorblockN{Shashwat Kumar}
\IEEEauthorblockA{Department of Engineering\\Systems and Environment\\
University of Virginia\\
Charlottesville,\\ Virginia, 22903\\
Email: sk9epp@virginia.edu@virginia.edu}
\and
\IEEEauthorblockN{Mehdi Boukhechba}
\IEEEauthorblockA{Department of Engineering\\Systems and Environment\\
University of Virginia\\
Charlottesville,\\ Virginia, 22903\\
Email: mob3f@virginia.edu}
\and
\IEEEauthorblockN{Arsalan Heydarian}
\IEEEauthorblockA{Department of Engineering\\Systems and Environment\\
University of Virginia\\
Charlottesville,\\ Virginia, 22903\\
Email: ah6rx@virginia.edu}}


%Telephone: (800) 555--1212, Fax: (888) 555--1212}


\maketitle

\begin{abstract}
Integrating driver, in-cabin, and outside environment's contextual cues into the vehicle's decision making is the centerpiece of semi-automated vehicle safety. Multiple systems have been developed for providing context to the vehicle, which often rely on video streams capturing drivers' physical and environmental states. While video streams are a rich source of information, their ability in providing context can be challenging in certain situations, such as low illuminance environments (e.g., night driving), and they are highly privacy-intrusive. In this study, we leverage passive sensing through smartwatches for classifying elements of driving context. Specifically, through using the data collected from 15 participants in a naturalistic driving study, and by using multiple machine learning algorithms such as random forest, we classify driver's activities (e.g., using phone and eating), outside events (e.g., passing intersection and changing lane), and outside road attributes (e.g., driving in a city versus a highway) with an average F1 score of 94.55, 98.27, and 97.86 \% respectively, through 10-fold cross-validation. Our results show the applicability of multimodal data retrieved through smart wearable devices in providing context in real-world driving scenarios and pave the way for a better shared autonomy and privacy-aware driving data-collection, analysis, and feedback for future autonomous vehicles.          
\end{abstract}


% \fi
\IEEEpeerreviewmaketitle

\begin{comment}

Outline of the paper:

1.Intro:
a.Semi-autonomous vehicle and the problem of shared autonomy
b.0	Cameras are great but…issues:
b.1.	Privacy is a good angle 
b.2.	Lighting condition
b.3.	Multiple cameras needed
c.passive sensing is a solution
d.	In this research…

2.Background:
a.Activity recognition using wearable this is our only focus nothing else. 

3.Method:
a.	Data collection
b.	Data processing 

4.Results:
a.	Showing category differences
b.	Predicting using either random forets or extra decision trees
c.	Loss and accuracy
d.	Confusion matrix
e.	Optional oversampling (SMOTE)

5.Discussion:
a.	Privacy
b.	Autonomous vehicle using 
c.	Don’t store raw readings of accelerometer
d.	Number of participants and data
\end{comment}

\section{Introduction}
Autonomous technologies are evolving exponentially, changing the interaction between semi-automated vehicles and the human driver drastically \cite{alsaid2020moving,hancock2019future}. Previously, advanced driving assisting systems (ADAS) (e.g., cruise control, or lane-keeping assisting system) were only used to assist the driver who acted as a supervisor; however, with the fast-paced improvements in auto-pilot systems (e.g., autopilot in TESLA vehicles), the line between the supervisor and a sole driver is starting to blur. This approach to autonomy is referred to as shared-autonomy, in which the task of driving is achieved through a collaboration between the driver and the vehicle \cite{fridman2018human}. Despite their successes, these approaches still rely on the driver to take over in critical situations without the vehicle having any understanding of the current state of the driver (e.g., distracted, fatigued, stressed, or under the influence of substances). In order to address these issues, the shared autonomy approach generally includes a driver sensing module in which the driver's states and behaviors are monitored in real-time. An example of a commercial semi-automated system that has extensive driver monitoring is the Comma.ai's \textit{openpilot} \cite{commaai} which monitors the driver in real-time and disengages the autonomy if the driver appears to be distracted. Through the driver sensing module, the vehicle is provided with a contextual awareness with respect to the driver's status in real-time. 

Driver sensing systems often rely on camera streams as they are very powerful in revealing driver's activities, states, and behaviors in real-time through different activity recognition \cite{pandriver,li2020detection}, distraction detection \cite{khurana2020eyes}, cognitive load estimation \cite{fridman2018cognitive}, and emotion detection methods \cite{abdic2016driver}. However, camera streams suffer from a few hardware, driving, and privacy-related issues. First, changes in illumination (e.g., night condition), having shadows, angle of the camera, quality of videos, and even type of the camera can significantly influence their accuracy and reliability in detecting driver's state and behaviors \cite{jegham2020vision,pandriver}. Second, to cover the entire in-cabin space, multiple cameras have to be used because one camera cannot cover all in-cabin and outside conditions \cite{roitberg2020open,ohn2014head}. Third, cameras are highly intrusive in nature, which may make people feel uncomfortable when being constantly monitored \cite{stark2020don}. Due to privacy concerns, many times, people prefer not sharing their video/audio data, making it impossible to track their states \cite{alharbi2018can}. Lastly, although we have seen significant improvements in computational techniques, processing video streams and applying computer vision techniques to detect human activity and state are still relatively costly and resource extensive \cite{hussain2020review}.

Although recent improvements such as enhancements in night vision capability of cameras or fusing radar systems with camera streams have overcome some of the aforementioned problems, the issue with privacy intrusion, cost, and high computational requirement still remain as main issues with cameras. An alternative solution is to use passive sensing techniques through smart wearable devices to detect driver behavior and states. These passive sensing methods can be used to (1) optimize video recording duration to the least amount possible (e.g., only collecting videos when passive sensing requires additional contextual information on the driver and environment); (2) cover situations that vision systems cannot capture (e.g., in low-light conditions); and (3) use less computation energy and cost as compared to video recording. While passive sensing has been studied in driving research, their applicability in real-world driving scenarios remains an open question, partly because these technologies are just starting to become popular and emerge as a long-term deployment solution \cite{Tavakoli2021}.   

In this study, we leverage ubiquitous sensing for classifying various elements of driving context, including driver's activity, outside events, and driving environment attributes in a fully naturalistic driving study. We first provide a summary of previous works in the driver's state and behavior recognition using wearable devices. Then we delve into our methodology for collecting and processing passive driver sensing data. We compare and contrast multiple machine learning methods on classifying driver's activities, outside events, and road attributes from non-intrusive data in real-world driving situations. Our work facilitates the transition to using off-the-shelf wearable devices for detecting driver's state and behaviors.

\section{Background Study}
Recent studies have provided insights into the application of wearable devices and ubiquitous computing in driving studies. These studies primarily focused on various states of the driver such as stress \cite{choi2017wearable,tavakolipersonalized}, drowsiness \cite{kundinger2020feasibility,choi2017wearable}, distraction \cite{bi2019safewatch}, fatigue \cite{choi2017wearable}, as well as different driving behaviors such as take-over readiness \cite{pakdamanian2020toward,pakdamanian2020deeptake,alrefaie2019heart}, driving maneuvers (e.g., turning) \cite{liu2015toward}, turning and lane changes \cite{huang2019magtrack}.

\cite{liu2015toward} used a combination of phone and wearable device to detect driving maneuvers such as steering wheel turning angle. In their study, authors first detect driver's sensor reading from the pool of driver, passenger, and vehicle sensor data with an accuracy of 98.9 \%. Then, based on the driver's hand inertial measurement unit (IMU) measurements, the steering wheel angle is estimated in a stationary vehicle. \cite{saeed2017deep} analyzed driver's arousal level in a driving simulator using a wearable device, worn on the participants' writs. In their study, authors used deep neural networks to detect three levels of arousal, under, normal, and over-aroused based on the driver's heart rate, skin conductance, and skin temperature. The authors were able to achieve an F-score of 0.82 for arousal detection through recruiting 11 participants. Similarly, \cite{choi2017wearable} used a wrist-worn wearable device for detecting driver's stress, fatigue, and drowsiness. By using a driving simulator with 28 participants, they were able to achieve an accuracy level of 98.43 \% for detecting each state. \cite{goel2018detecting} used a wearable smartwatch to detect instances of driver's distraction in a driving simulator. In their study, by collecting physiological data from 16 participants and utilizing multiple machine learning algorithms such as decision trees and support vector machines to classify different driving states, they achieved an average accuracy of 89 \% in distraction detection. 

\cite{bi2019safewatch} used a wearable device and a phone for detecting driver's distraction based on driver's hand movement. In their study, authors developed an app that detects various motions of the driver's hands, such as different types of holding the steering wheel, by fusing the sensor readings from both driver's hand and the phone's IMU measurements. By using the IMU sensors on the wearable device  \cite{li2019recognition} was able to detect different classes of phone, text, drink, using a marker, using a touch screen, and normal driving with an F-1 score of 0.87. \cite{huang2019magtrack} used a smartwatch and wearable magnetic ring along with eye-tracking glasses to detect driver's distraction and activities such as turning and lane changes. Their systems achieved 87 \% precision and 90 \% recall on data collected from 10 participants in an on-road controlled study. In a recent study, \cite{kundinger2020feasibility} used off-the-shelf wearable devices to detect driver's drowsiness in a driving simulator. In their study, authors have used heart rate signals for drowsiness detection and achieved an accuracy of 99.9 \%. This study also pointed out the differences between age groups for drowsiness detection. \cite{tavakolipersonalized} used wearable data from 12 participants and compared drivers' heart rate variability in different road types (i.e., city versus highway), weather conditions (i.e., rainy versus clear), and presence of a passenger. They found that drivers are on average calmer on highways, in clear weather, and when being accompanied by a passenger.  

Although these studies provided rich insights into the application of wearable devices in driving, they are mostly performed in controlled environments (i.e., driving simulators and on-road controlled studies). This can become problematic because most of these studies lack the proper ecological and external validity to allow their findings to generalize in real-life contextual settings where different real-world challenges and noise exist. Additionally, most of these studies have only leveraged the IMU sensor (i.e., accelerometer and gyroscope) on the smart wearable for driver behavior detection leaving other sensors such as heart rate (HR) and Photoplethysmography (PPG) unexplored. Furthermore, as opposed to most of the studies that used wearable devices that were designed specifically for the study of interest, we are interested in exploring ubiquitous sensors such as off-the-shelf devices (e.g., smartwatches) for driver state and behavior analysis in the wild. This is due to the fact that these devices are (1) already used by drivers and participants in their daily lives, and (2) they have multiple sensors built into them.

\section{Methodology}

We now present our approach. We first describe the data collection and annotation protocol. Then, we provide details on the feature exploration to test the feasibility of the classification problem. Finally, we describe our feature extraction and machine learning classification pipeline.

\subsection{Data Collection}
The data in this study has been collected through HARMONY, a human-centered naturalistic driving study platform \cite{Tavakoli2021}. Here we provide an overview of the data collection and cleaning process. However, the reader is invited to refer to \cite{Tavakoli2021} for more detailed information on the data collection, storage, and processing within the HARMONY system. A sample data of HARMONY is made publicly available to researchers and can be retrieved from \cite{Harmonydata}. 

HARMONY is a framework that is leveraged to collect driver's behavioral and state measures together with environmental sensing, through physical devices such as cameras (both inside and outside), and wearable devices, as well as virtual sensors such as computer vision algorithms or APIs to extract features from the data collected by the physical sensors. Within the current HARMONY dataset and framework, 21 participants' vehicles are equipped with a BlackVue DR-750S-2CH dash camera, which records both inside and outside of the vehicle as the car turns on and the driving scenario starts. Each driver is provided with an android smartwatch, equipped with an in-house app, namely SWear \cite{boukhechba2020swear}. SWear is available on Google Play store \cite{boukhechba} and is designed to facilitate the process of data collection on smartwatches by adding the ability to control each sensor's data collection frequency to the desired sensing specifics and sync sensors data to the cloud. Through this, we collected drivers' hand acceleration [10Hz], gyroscope measures [10Hz], Photoplethysmography (PPG) [10Hz], ambient audio amplitude (noise level) [1/60 HZ], HR [1Hz], ambient light intensity [1/60 HZ], and GPS location [1/600 HZ]. The data collection set-up in participants vehicle can be viewed on Fig. \ref{fig:equipment} - A.

\subsection{Data Annotation and inspection}
\begin{figure*}
  \centering
  \includegraphics[width=1\linewidth]{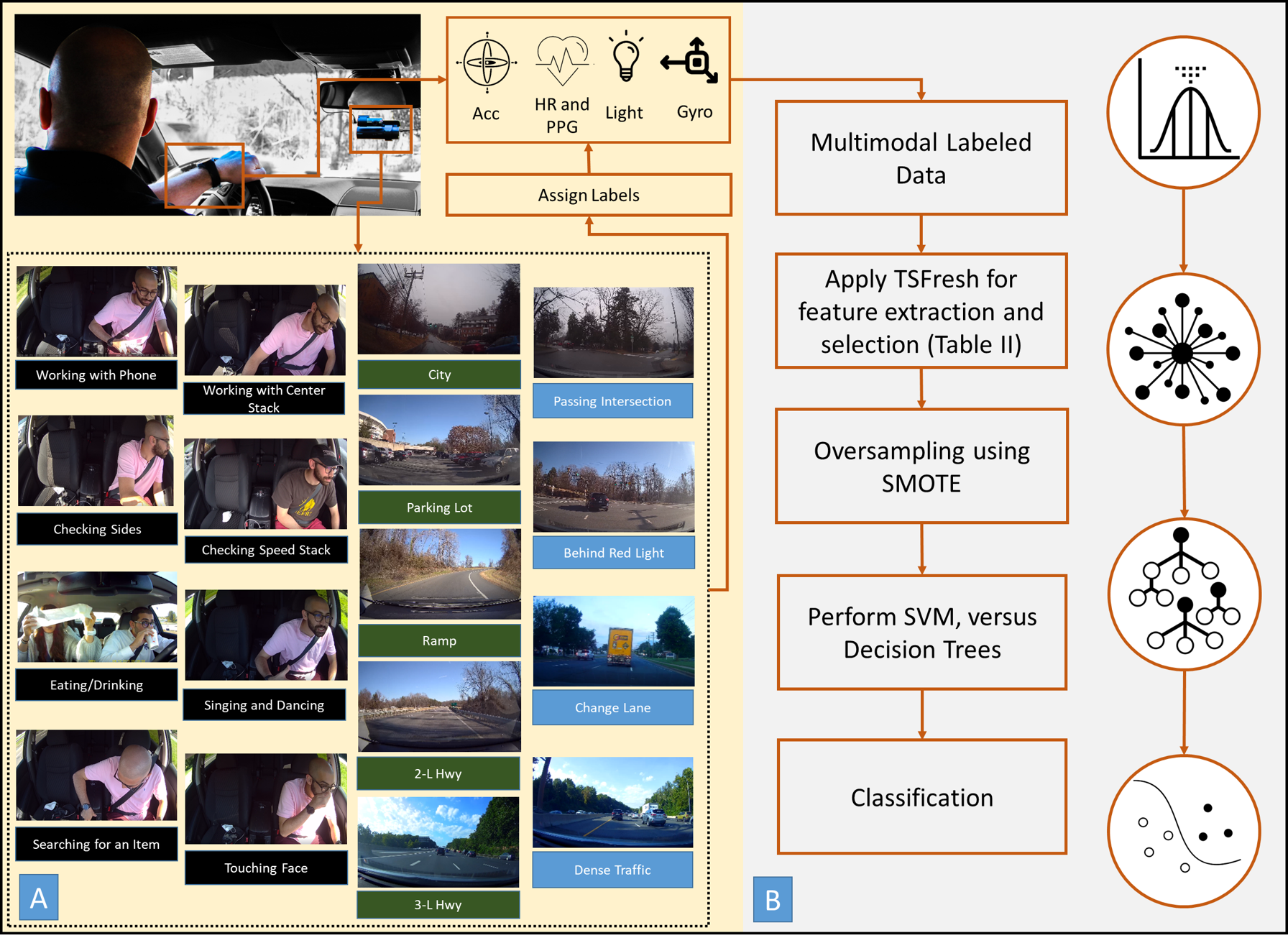}
  \caption{Summary of our methodology for integrating smartwatches into driver behavior analysis in the wild. We use videos to label smartwatch data}
  \label{fig:equipment}
\end{figure*}

We have chosen a random subset of the collected data for the purpose of this study from 15 participants within HARMONY. We use the in-cabin and outside facing videos for manual annotation of driver's activities, outside events, and environmental attributes for training purpose (Fig. \ref{fig:equipment} - A). For this study, we have divided up driver activities into working with the mobile phone, checking sides, which often includes checking mirrors, checking speed stack, checking center stack, eating/drinking food, searching for an item on the passenger's seat, touching face, and singing and dancing. Outside events consist of passing an intersection, staying behind a traffic light, being stuck in dense traffic, and changing lanes. The road conditions are divided into the city streets, 2-lane highway, 3-lane (or more) highway, parking lot, and merging ramps. Since all devices' timestamps are synched, we use the start time of these events to pull the wearable data, such as IMU, PPG, light, and noise levels. A sample of annotated data can be view on Fig. \ref{fig:equipment} - A. Table \ref{tab:data} shows the final count of data samples per activity as used for this study. 

\begin{table}[]
\caption{Summary of data samples used for classification}
\label{tab:data}
\resizebox{0.49\textwidth}{!}{%
\begin{tabular}{|cccc|}
\hline
\multicolumn{1}{|c|}{\textbf{Category}} &
  \multicolumn{1}{c|}{\textbf{Event}} &
  \multicolumn{1}{c|}{\textbf{\begin{tabular}[c]{@{}c@{}}No. of \\ Samples\end{tabular}}} &
  \textbf{\begin{tabular}[c]{@{}c@{}}Average \\ Duration\\ (sec)\end{tabular}} \\ \hline
Inside Activity & Checking Sides               & 128000 & 1   \\
Inside Activity & Eating/Drinking              & 27584  & 4   \\
Inside Activity & Working with Center Stack    & 6540   & 3.5 \\
Inside Activity & Checking Speed Stack         & 24000  & 1   \\
Inside Activity & Touching Face                & 14500  & 5.3 \\
Inside Activity & Working with Phone           & 47000  & 22  \\
Inside Activity & Singing and Dancing          & 5870   & 22  \\
Inside Activity & Searching for an Item        & 2680   & 10  \\ \hline
Outside Event   & Change Lane                  & 22437  & 3   \\
Outside Event   & Passing an Intersection & 36722  & 3   \\
Outside Event   & Traffic Light                & 42670  & 20  \\
Outside Event   & Stuck in Traffic             & 4542   & 60  \\ \hline
Road Type       & City Street                  & 182614 & 158 \\
Road Type       & Parking Lot                  & 12553  & 38  \\
Road Type       & Merging Ramp                 & 4984   & 26  \\
Road Type       & 2L - Highway                 & 180295 & 168 \\
Road Type       & 3L - Highway                 & 104102 & 132 \\ \hline
\end{tabular}%
}
\end{table}

\begin{figure}
  \centering
  \includegraphics[width=1\linewidth]{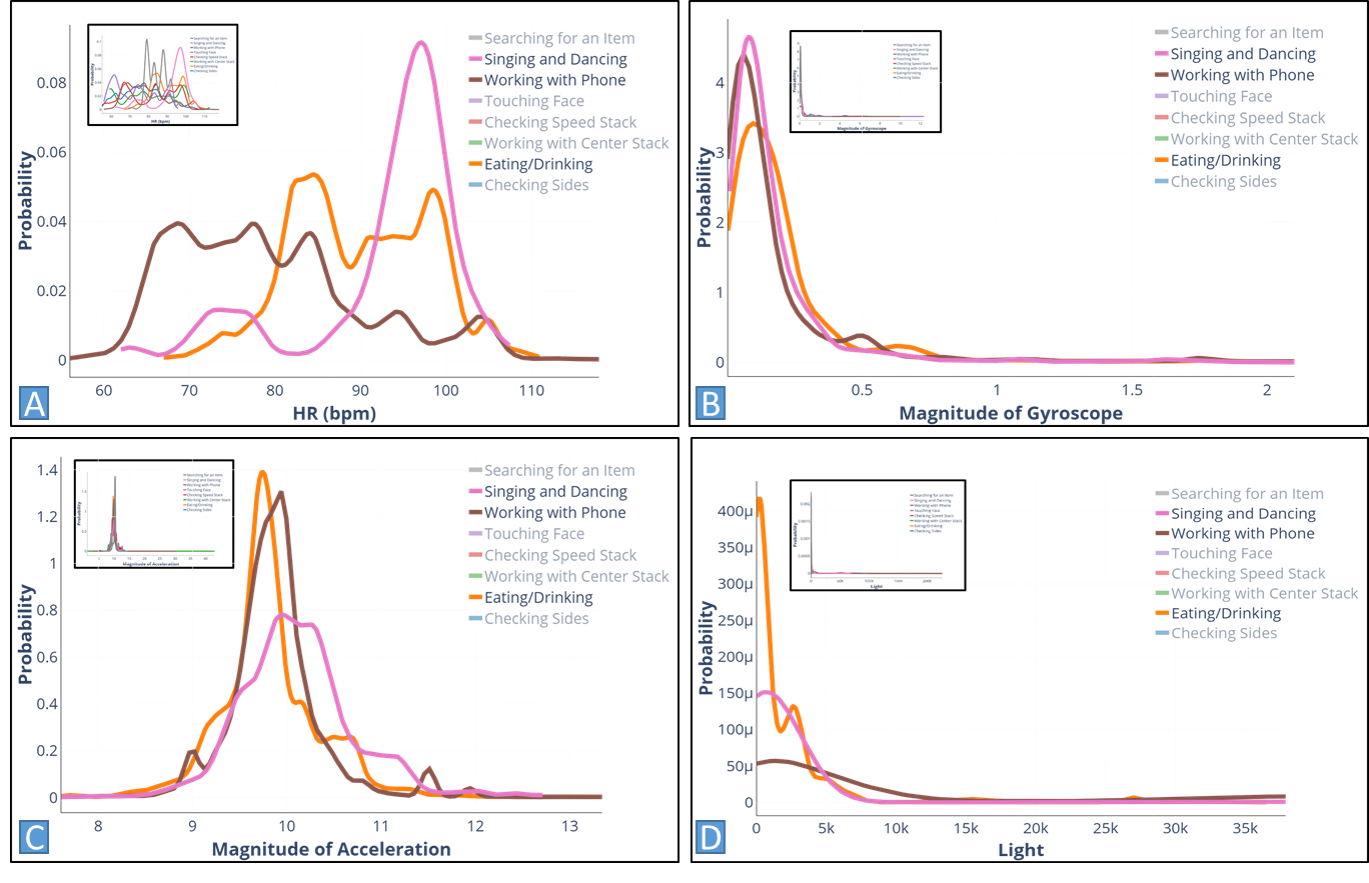}
  \caption{Distribution of HR (A), accelerometer (B), gyroscope (C), and light (D) data among different activities. This example shows how the distribution of each modality varies across activities and motivates its use as a predictor of drivers states. Note the distribution differences among different activities, such as shifted HR peaks for singing as compared to food-related and phone-related activities.}
  \label{fig:hr}
\end{figure}

Additionally, we inspected the aforementioned categories prior to classification for visual observation of differences in sensor reading distribution. This helps investigate how sensing data fluctuates across these settings and lay the ground for the feasibility of classifying them through machine learning algorithms. For instance, drivers' heart rate follows different distributions when performing different activities. This can be seen on Fig. \ref{fig:hr} - A where the distribution of heart rate in singing has a peak at a higher heart rate level as compared to the food-related (i.e., eating and drinking) category. Additionally, we observe differences in the distribution of IMU and light sensors as depicted on Fig. \ref{fig:hr}.   

\subsection{Feature Extraction}
Working with the data extracted from commercially available wearable devices has two main challenges; First, The average data collection frequency of each modality can be different. Additionally, each sensor's data collection frequency is not constant, and it may vary throughout the time due to hardware requirements (e.g., the frequency of HR data varies between 0.8Hz to 1.1 Hz at different battery levels). Although we can change the average frequency value through SWear, the exact frequency cannot be defined prior to data collection. To solve these challenges, we resampled the data at 10 Hz, which is equal to the IMU, Light, and PPG sensor average frequency, which have the lowest sampling rate. 

In order to perform feature extraction, we analyzed the duration of different activities to find the correct sliding window. Additionally, previous literature (\cite{twomey2018comprehensive}) on accelerometer-based activity recognition had stated a window overlap of 50 percent to avoid correlations between samples. We chose a sliding window of 1 second to cover all the activities duration (see Table \ref{tab:data}) in our feature extraction, with 50 \% overlap. In order to perform feature extraction, we leveraged the tsfresh package \cite{christ2018time} on Python to retrieve multiple time and frequency domain features from each individual wearable sensor (i.e., Light, PPG, HR, and IMU). Additionally, we have leveraged previous literature in human and driver activity and state recognition using biosignals to select relevant features from the tsfresh library \cite{attal2015physical,twomey2018comprehensive,pakdamanian2020deeptake,hernandez2020literature,sousa2019human,sousa2019human}. Table \ref{tab:feats} shows the selected features for this study.  

% \usepackage{graphicx}
\begin{table}[]
\caption{Summary of selected features}
\label{tab:feats}
\resizebox{0.49\textwidth}{!}{%
\begin{tabular}{cc}
\hline
\textbf{Domain} &
  \textbf{Features} \\ \hline
Time &
  \begin{tabular}[c]{@{}c@{}}Kurtosis, mean, standard deviation, maximum, minimum, \\ variance, skewness, median, variation coefficient, absolute sum of changes, \\ Benford correlation, count above mean, count below mean, \\ first location of maximum, first location of minimum,\\ has duplicate, has duplicate max, has duplicate min, \\ last location of maximum, last location of minimum, \\ longest strike above mean, longest strike below mean,\\ mean abs change, mean change, mean second derivative central,\\ sum of reoccurring data points, sum of reoccurring values, sum values\end{tabular} \\ \hline
Frequency &
  Energy, power, entropy \\ \hline
\end{tabular}%
}
\end{table}

Because the frequency of different activities varies across different driving scenarios, the dataset has an unbalanced nature with respect to different classes. For instance, in our annotated dataset, the number of the mirror checking events is noticeably higher than the phone usage activities (see Table \ref{tab:data}). To solve this issue, we have used two separate methods of: 
\begin{enumerate}[label=(\Alph*)]
    \item balancing the classes by weighing each class to be inversely proportional to its frequency in the dataset. 
    \item Oversampling based on Synthetic Minority Oversampling Technique (SMOTE) \cite{chawla2002smote} to generate new samples for the minority classes. SMOTE generates new data points from convex combinations of nearest neighbours. We apply SMOTE only on the training set. This is important for keeping the classifier unbiased towards new samples. 
\end{enumerate}

Using the extracted features, We built different models to classify three major groups of driver's activities, environmental events, and environmental attributes as described earlier. In each category, three different machine learning methods were used to build classifiers using a 10-fold cross-validation method. The classifiers included random forest, decision trees, and extra decision trees. We observed that the Random Forest model outperforms the other two classifiers, thus we only focus on providing the results for the Random Forest classifier. For this, the Scikit-learn package \cite{scikit-learn} was utilized. 

\section{Results}
\begin{figure}
  \centering
  \includegraphics[width=1\linewidth]{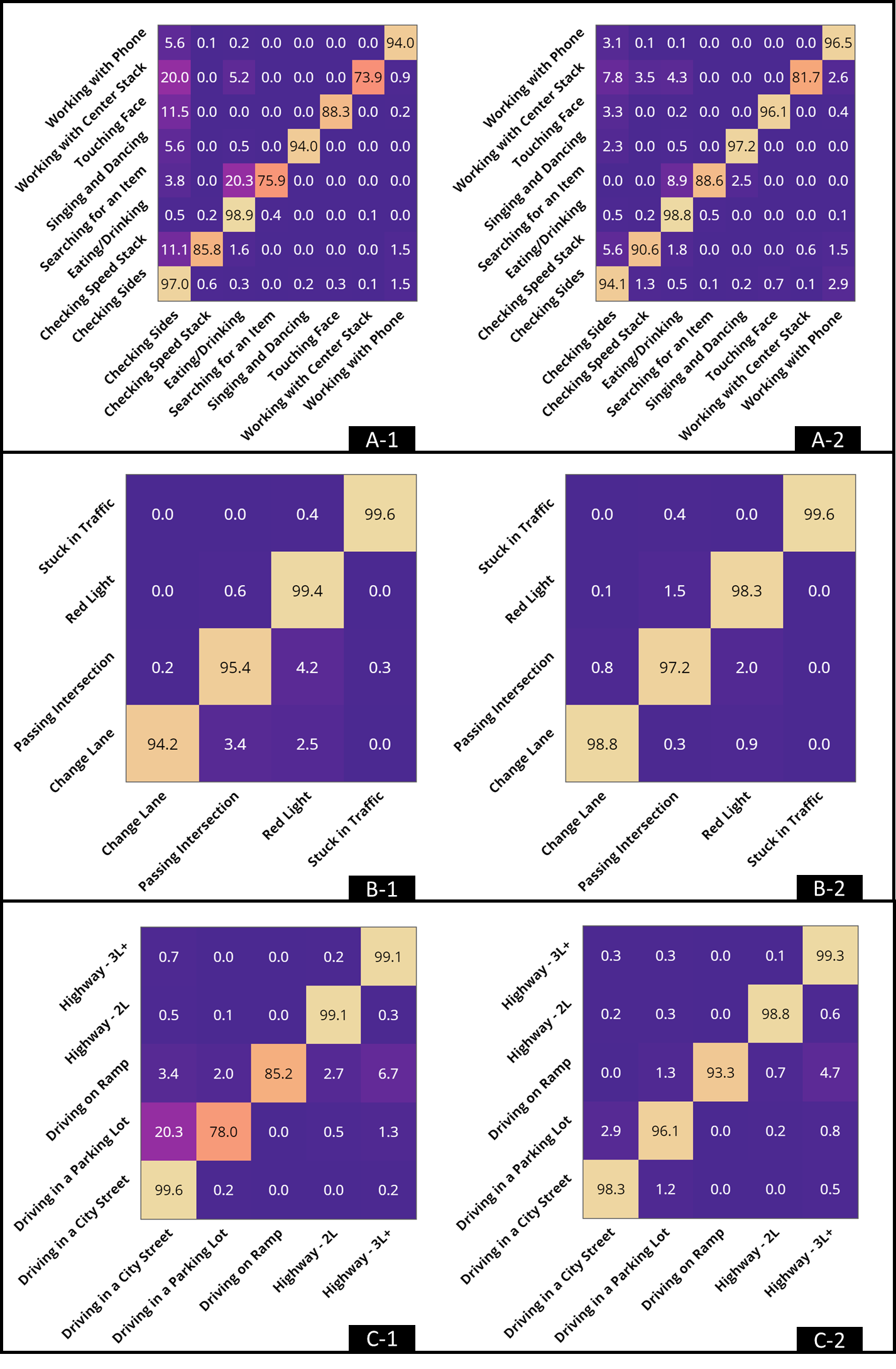}
  \caption{The confusion matrix for classifying driver's activity (A), outside events (B), and road type (C), in two categories of unbalanced data (1) and oversmapled data (2)}
  \label{fig:conf}
\end{figure}

Fig. \ref{fig:conf} demonstrates the confusion matrix for each of the three classification tasks (A: driver activity, B: outside events, and C: road types) using the highest accuracy model trained on the original data, as well as the models trained on oversampled data (e.g., A1 is the original data and A-2 shows the oversampled data). 

For classifying activities, the Random Forest model outperforms the other two methods (i.e., decision trees and extra decision tress), with an average F1 Score of 90.99 \% on the imbalanced data (\ref{tab:results}). This model does relatively poorly on the two categories of ``searching for an item," and ``working with the center stack display of the vehicle" (Fig. \ref{fig:conf} - A-1 and A-2). The searching for an item category is mostly due to lower amount available data. The center stack category may be mistaken with the checking sides category 20 percent of the time. We suspect this is mostly due to the fact that the participants body movements are very similar when working with the center stack and checking different sides of the vehicle (e.g., checking side mirrors). To further confirm this, we perform the SMOTE oversampling on the training set and train the classifier on the oversampling data. It is recognizable that the accuracy of detecting the searching for an item category increases significantly (i.e., from 75.9 \% to 88.6.3 \%), while the center stack detection accuracy does not increase as much (i.e., from 73.9 \% to 81.7 \%) (Fig. \ref{fig:conf} - A-1 and A-2). Finally, the model trained on oversampled data, on average achieves an F1 score of 94.55 \%. 

For classifying outside events, the Random Forest classifier outperforms the other methods (Table \ref{tab:results}) with an average F1 score of 97.68 \%. Using SMOTE, we can further enhance the classification F1 score to 98.26 \%. Oversampling, mostly enhances change lane and passing intersection categories but does not change the overall accuracy significantly. Similarly, Random Forest outperforms other models in detecting the road type with an average F1 score of 93.62 \% and 97.68 \% for the original data and oversampled data respectively. Due to lower amount of data, the model performs relatively poorly on the Driving in a Parking Lot and Driving on Ramp categories, which is then further enhanced by using the oversampling methods.

\begin{table}[h]
 \caption{Summary of model performance on different target categories when applied with a 10-fold cross validation. Note the enhancements in performance when generating new samples using the SMOTE method.}
    \label{tab:results}
\centering
  \begin{tabular}{c|c|c|c|c}
     & \multicolumn{2}{c}{Imbalanced}  & \multicolumn{2}{c}{SMOTE} \\

    \hline
       Target Category & F1 score & SD & F1 score & SD \\

     \hline
     Driver Activity & 90.99 & 0.92 & 94.55 & 0.77  \\
     Outside Events & 97.68 & 0.69 & 98.27 & 0.28  \\
     Road Type & 93.62 & 0.69 & 97.86 & 0.25  \\

    \end{tabular}
 
\end{table}

\begin{figure}
  \centering
  \includegraphics[width=1\linewidth]{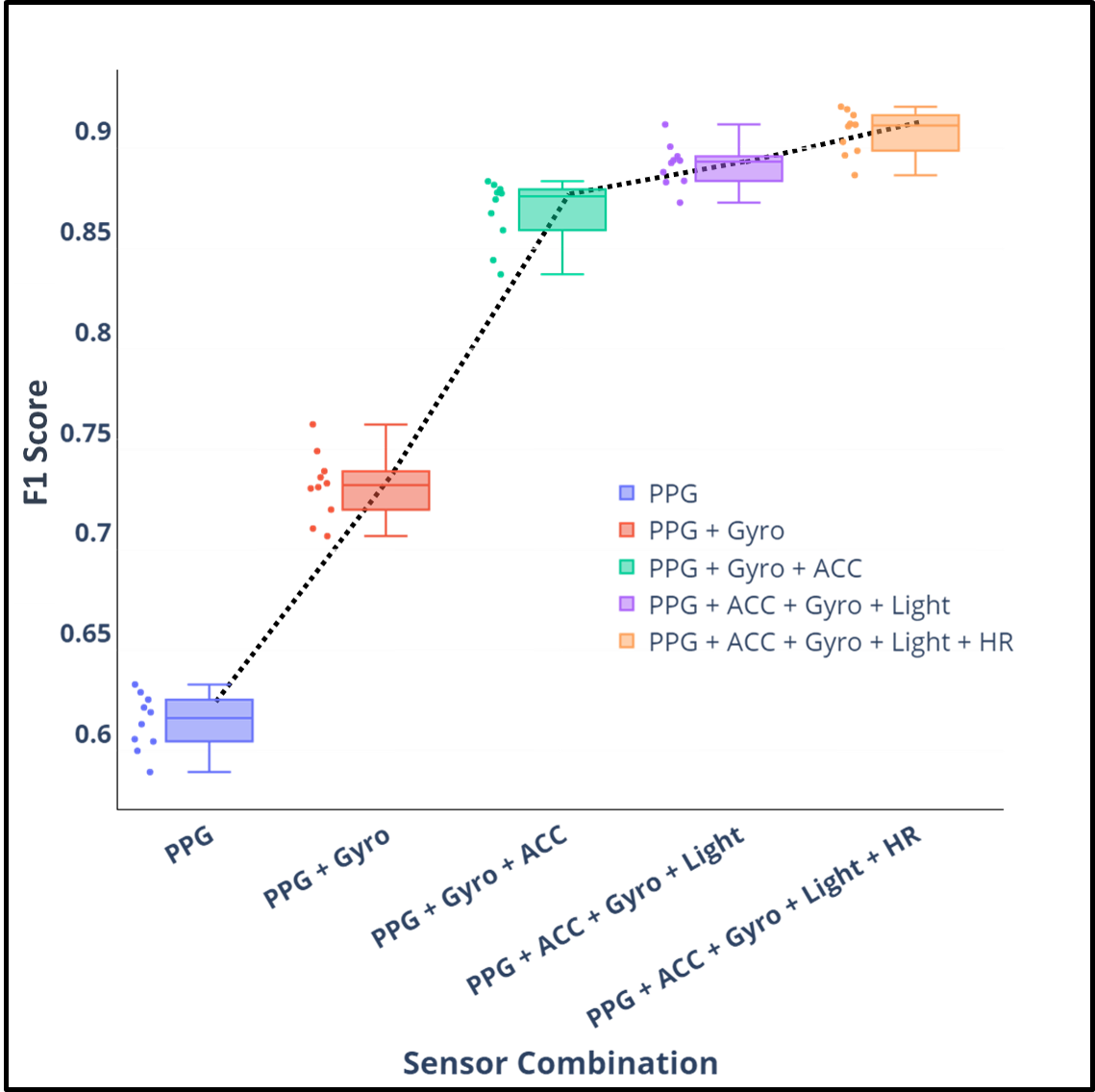}
  \caption{The comparison of adding each sensor to the classification scheme in a 10 fold cross validation scheme for driver activity classification model. Note the increase in F1-Score as new modalities provide additional information to the classifier.}
  \label{fig:summ}
\end{figure}

Additionally, we assessed the contribution of multimodality to the classification accuracy of the driver's activity model. We emphasize on this attribute of current off-the-shelf smart wearable devices as previous works were mostly focused on laboratory-built devices, lacking multimodal sensor data built into one device. Fig. \ref{fig:summ} demonstrates the accuracy of the activity recognition model trained on the original data using Random Forest classifier with varying sensor combinations. As shown, overall by adding different modalities, the activity recognition model's F1 Score increases. 

\begin{figure*}
  \centering
  \includegraphics[width=1\linewidth]{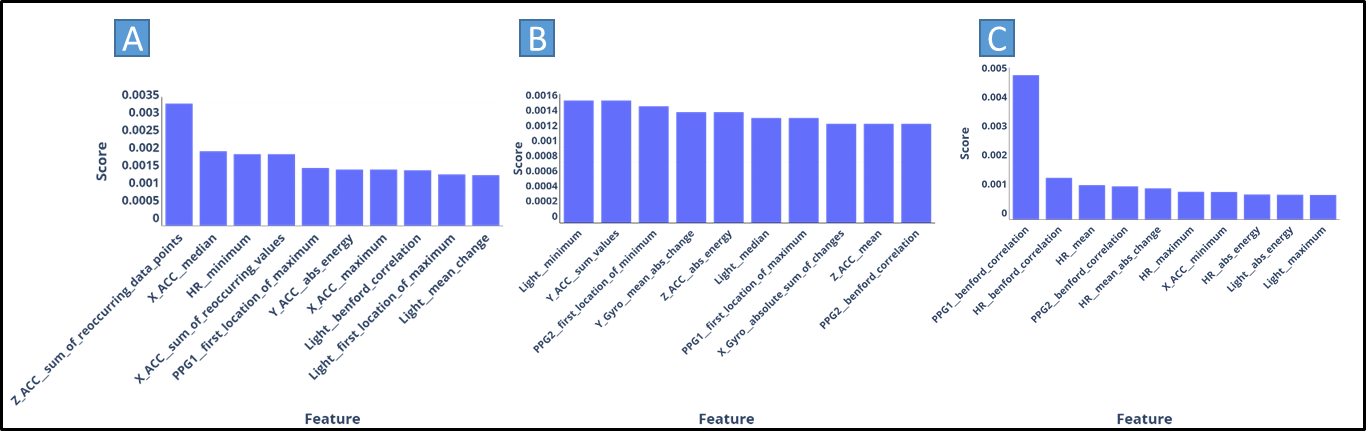}
  \caption{Feature importance analysis on the three targeted categories of  (A) driver's activities, (B) outside events , and (C) road types. Note that the contribution of different modalities varies across the different targeted predictions.}
  \label{fig:featimp}
\end{figure*}

To further analyze the effect of each modality on the classification task, we performed a permutation feature importance analysis \cite{scikit-learn} on the three aforementioned classifiers. For simplicity, this analysis is performed only on a hold-out sample in our dataset. Similar to the previous section, the training set is oversampled through SMOTE. Fig. \ref{fig:featimp} shows the top ten features in each target category of driver activity, outside events, and road type. As shown on Fig. \ref{fig:featimp} - A for classifying the driver's activities, five out of the top ten features are from the accelerometer modality, which draws attention to the importance of the IMU sensor for activity recognition similar to the previous research as mentioned in the background section. However, as shown on Fig \ref{fig:featimp} - B, heart rate, PPG, and light are also contributing to the classification task, pointing out the importance of multimodal sensing. Additionally, for classifying the outside events, we observe the inclusion of gyroscope in the top ten features, which points out the differences among driver's hand rotations for certain outside events (e.g., change lane action). Moreover, note that three out of the top ten features are related to the PPG sensor, which is shown to be correlated with driver's state, such as stress levels \cite{kim2018stress}. Lastly, the PPG sensor is contributing the most to the road type detection task. This is in line with previous research demonstrating the variation in driver's HR when driving in different road and weather conditions \cite{Tavakoli2021}.               

\section{Discussion}
%we provided a context to the vehicle while being low cost and privacy friendly
In this study, we have classified real-world driving instances such as driver's activities, outside events, and outside road attributes using data extracted from commercially available smartwatches. Smartwatch multimodal data has the potential to reveal different aspects of driving behavior and activities such as movements (e.g., through IMU sensors), underlying state of the driver such as stress (e.g., through HR sensor), and environmental attributes (e.g., through noise and light sensors). Note that with the current developments in ubiquitous computing devices, multiple other sensors such as skin temperature, breathing patterns, skin conductance, and even electrocardiogram (ECG) are starting to be built into these devices, which can further push the boundaries of driver state recognition. While one sensing modality (e.g., Light) might not be able to provide all the information to detect certain activities, other sensing modalities such as IMU may still provide crucial missing information to aid the classifier (see Fig. \ref{fig:summ}). As a result, by having access to different modalities, we can identify different activities and events with higher accuracy. This has also been shown in our previous work \cite{Tavakoli2021} that certain environmental events might leave a driver's visual and behavioral patterns unchanged, while affecting his/her underlying state, which can be measured through HR. This highlights the importance of multimodal sensing when performing classification in real-world scenarios. 

%you need to have some discussion the data and the results you have. you are able to detect these categories and activities with this level of accuracy across 21 participants data. explain what is novel about this work and how this can help future shared autonomy discussions you had
% explain the novelty/benefit of your methodology; also highlight what can be tested and expanded. e.g., we can test more advanced models on these datasets...(addressed in first and last paragraph)
% talk about what are things that can only be seen through having access to naturalistic dataset...highlight that for what you are proposing, you need to have naturalistic dataset...can you show that with what you have?
% so you were able to classify these events through this technique that you have, then what? (this is in the privacy section paragraph below) 

Our system can be integrated with cameras and other vision systems (e.g., eye trackers) to provide more accurate detection of driver's activities and behaviors. Previous studies have developed deep-learning based methods for classifying driver's activities, and behaviors, as well as outdoor events using video cameras when the lighting conditions are suitable \cite{pandriver,li2020detection,fridman2018cognitive,khurana2020eyes}. The integration of ubiquitous computing in such systems through sensor fusion techniques can help in edge cases of vision system applications (e.g., low illumination condition) and in cases that the video streams are not available due to different reasons (e.g., driver not allowing for video recording, lack of visibility due to lighting). For example, by integrating smartwatches into vehicles, a user can request between two levels of privacy. In a high privacy case, driver's activities, and behaviors are monitored through the smartwatch. In the other case, the user can choose to provide more data streams, in which the driver's video can be fused with the smart wearable to detect different activities, behaviors, and environmental attributes. In the case of low privacy mode, depending on the light level measured through an illuminance sensor (e.g., smartwatches often provide this sensor), the system can perform using both camera and smartwatch or by only relying on the smartwatch. All of these decisions and analysis can also be performed using edge computing devices that are already being used for autonomous vehicle development and data collection purposes such as NVidia Jetson family, which have onboard GPU and other computing units for real-time analysis. Additionally, it should be considered that wearable data is much smaller than videos and requires lesser computational resources as compared to bigger video-based deep learning based models. 

Our work can be improved from multiple aspects. First, by labeling more driving data, we can enhance the accuracy of our method for driver activity classification. Although we are using manual annotation for providing ground truths, we can optimize our labeling speed by using more automated approaches such as leveraging change point detection methods \cite{Tavakoli2021}. By having more labeled data, we can then use other classification methods that require a higher amount of data (e.g., deep neural networks). Additionally, we will build and implement a system that can perform the modeling scheme provided here either on the watch, using an embedded system (e.g., Nvidia Jetson) in the vehicle, or on the cloud, to optimize our system for real-time applications (e.g., battery considerations). This can also help us make our system more human-centered by incorporating user feedback. Lastly, we will continue to test newer wearable devices to increase our number of sensor modalities (e.g., adding skin temperature), which can further increase the accuracy of our system.

% you need to talk about limitations of the work -
%% there are limitations that we have manual training, we need more data, we need real-time computation...you need to expand on these here or in the next section

\section{Conclusion and Future Work}
In this study, we have analyzed driver's activities and behaviors through smartwatches in naturalistic conditions. In contrast to previous works, we have used real-world fully natural data through off-the-shelf sensing devices in driving scenarios. Through using different machine learning algorithms, we were able to accurately classify driver's activities, behaviors, and environmental attributes. Our study is taking three directions in the future. First, we will add more participants and more driving actions to diversify our classification. Second, we will be fusing smartwatch readings with the other modalities of data (e.g., driver's gaze directions) to run different experiments for analyzing the contribution of each modality (i.e., vision versus smart wearable) in driver state and behavior detection. Third, we will be testing our system with newer smartwatches to analyze the effect of adding new sensors (e.g., skin temperature, skin conductance, and breathing patterns) to the classification accuracy.


\section*{Acknowledgment}
We would like to thank the University of Virginia (UVA) Link Lab, and the Virginia Commonwealth Cyber Initiative (CCI) for providing support and resources to enable this project. Also, we are thankful to the UVA Institutional Review Board for their continuous support and feedback.



\bibliographystyle{IEEEtran}
\bibliography{IEEEexample}

\end{document}